\documentclass[pra, twocolumn,nofootinbib,
 superscriptaddress,
 showpacs,
 longbibliography,
 aps,
 showkeys]{revtex4-2}

\usepackage{lmodern}
\usepackage[T1]{fontenc}
\usepackage{graphicx} 
\usepackage{mathtools,color,amsmath}\usepackage[a4paper]{geometry}
\usepackage{amsfonts,amssymb,amsmath,subfigure,float}
\usepackage{hyperref}
\usepackage[utf8]{inputenc}

\begin{document}
\title{Generation of Large Kitten states via thermally driven dissipative freezing 
}

\date{\today}
\author{Caspar Groiseau}
\email{groiseau@chalmers.se}
\affiliation{Departamento de F\'isica Teórica de la Materia Condensada and Condensed Matter Physics Center (IFIMAC), Universidad Autónoma de Madrid, 28049 Madrid, Spain}
\affiliation{Department of Microtechnology and Nanoscience (MC2), Chalmers University of Technology, SE-412 96 Gothenburg, Sweden}

\begin{abstract}
We show that one can exploit the representation of the Dicke state with magnetic quantum number $0$ in orthogonal bases to increase the probability of producing a large Kitten state or even a Cat state via dissipative one-axis twisting, resulting in a net increase in metrological sensitivity and robustness against white noise with respect to the initial state. The resulting states are also useful for quantum secret sharing. The success of the scheme is heralded by the rate of photoemissions. Unlike previous iterations, we have an important population of the larger Kitten states,
which makes post-selecting trajectories for them feasible.

\end{abstract}
\maketitle

\section{Introduction}

A Cat state, named in analogy to Schr\"odinger's thought experiment \cite{schrodinger1935}, is a superposition of two quasi-classical states. They are susceptible even to small perturbations, as can be seen by their high frequency interference patterns in the Wigner function. This makes them excellent for quantum-enhanced measurements and important for the field of quantum metrology \cite{pezze2018}. The generalized Greenberger-Horne-Zeilinger (GHZ) state is a specific type of Cat state important in the context of quantum secret sharing \cite{hillery1999}.

Cat states have been realized in a great variety of systems ranging from optical photons \cite{sychev2017,ourjoumtsev2007}, single Rydberg atoms \cite{facon2016}, cold atoms \cite{chalopin2018}, microwave photons with superconducting circuits \cite{vlastakis2013}, trapped ions \cite{monroe1996} and SQUIDs \cite{friedman2000}. The Cat states based on the superposition of photonic states are relatively fragile \cite{glancy2008,zhang2013}. In that regard, the angular momentum degree of freedom of matter seems to be more promising in producing large Cat states. However, in spinor systems the time scales can be slow due to weak interaction between particles. Recent initiatives have addressed this by implementing effective Dicke models, known for their strong cavity-mediated collective interactions. This has been achieved in atoms \cite{zhiqiang2017}, trapped ions \cite{aedo2018,safavi-naini2018} and BECs \cite{baumann2010,baumann2011}.

Decoherence or dissipation, stemming from quantum systems being open and in constant interaction with an environment, represents one of the largest obstacles in modern quantum physics, as it typically destroys the coherences that make quantum dynamics so advantageous. It can, however, also be a constructive effect in certain circumstances, e.g., when trying to generate a quantum state that is a dark or steady state. Proposals to exploit dissipation include dissipative quantum computation \cite{verstraete2009}, stabilization of topological edge states \cite{wetter2023}, improvement of the purity and phase coherence of BECs \cite{witthaut2008}, and preparation of entangled \cite{marr2003,kraus2008,diehl2008} and spin-squeezed states \cite{dallatorre2013}.

\begin{figure}[tbh]
\centering
	\includegraphics[width=0.9\linewidth]{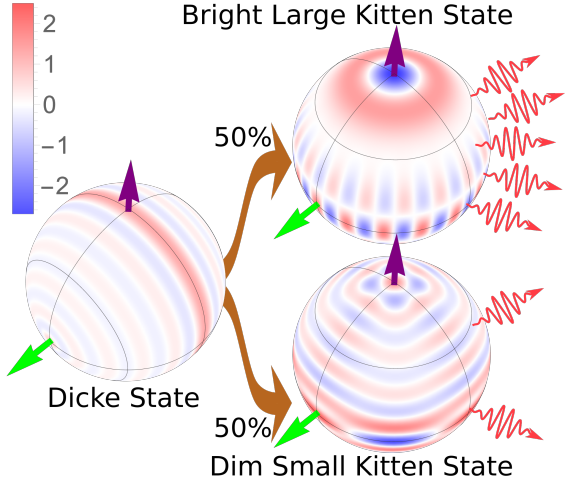}
	\caption{Schematic representation of the idea: the Dicke state $|S=10,m=0\rangle_x$ is transformed into either a large or a small Kitten state with equal probability. We display here the Wigner functions \cite{pezze2018,agarwal1981} of a large Kitten state exemplified by $(|S=10,m=9\rangle_y+|S=10,m=-9\rangle_y)/\sqrt{2}$ and a small Kitten state exemplified by $(|S=10,m=2\rangle_y+|S=10,m=-2\rangle_y)/\sqrt{2}$. The green arrow points in $x$-direction and the purple arrow in $y$-direction.}
	\label{idea}
\end{figure}

We build on the work in \cite{chia2008,groiseau2021,sanchezmunoz2019}, in which it was observed that a symmetric superposition of the eigenstates of the Liouvilian evolving under "dissipative one-axis twisting" will settle (or freeze) in an entangled-state cycle. These entangled-state cycles consisted of pairs of Kitten states (Cat state with smaller magnetic quantum number $|S,\pm m\rangle,{ }m<S$), between which the system indefinitely alternates via photoemissions. However, the likelihood of these cycles was determined by the initial population of the corresponding Dicke states. In \cite{chia2008,groiseau2021} a spin-coherent state (in a basis orthogonal to the twisting axis) and in \cite{sanchezmunoz2019} an equal superposition of all eigenstates were considered. The former had vanishing population of the large Kitten states and for the latter, it is unclear how to obtain it. Here we make use of the fact that the non-spin-coherent Dicke states, which can be experimentally achieved, have also representation as symmetric superpositions in orthogonal bases and show this can massively improve the likelihood of obtaining a large Kitten state. This is exemplified here for the optimal choice $|S,0\rangle$. FIG.~\ref{idea} shows the Wigner functions of this initial state and of resulting large and small Kitten states.

The paper is organized as follows: first, we present how to get from the Dicke model to the required dissipative one-axis twisting dynamics. Then, we show how the Dicke state $|S,0\rangle_x$ evolves under these dynamics and show how this can result in large Kitten states. 
Then we explore the potential usefulness of these states in quantum secret sharing and quantum metrology. 
Following that, we discuss the experimental feasibility, specifically the generation of the initial state and the influence of single-particle decoherence. Finally, we make our closing remarks and provide an outlook.

\section{Model}
We consider a system of $N$ identical spinors with total integer spin $S$ interacting with a cavity mode $\hat a$ in a Dicke model. The master equation for the density operator $\hat{\rho}$  for this configuration is given by $(\hbar=1)$
\begin{equation}
    \dot{\hat\rho}=-i\left[\hat H_\text{Dicke},\hat\rho\right]+\kappa(\mathfrak{n}+1) \mathcal{D}[\hat a]\hat\rho+\kappa\mathfrak{n} \mathcal{D}[\hat a^\dagger]\hat\rho,
\end{equation}
where we account for a thermal population of the cavity  $\mathfrak{n}$. We define the Lindblad superoperator $\mathcal{D}[\hat O]\hat\rho=2\hat O\hat\rho\hat O^\dagger-\hat\rho\hat O^\dagger\hat O-\hat O^\dagger\hat O\hat\rho$ \cite{breuer2002}. The Hamiltonian of the Dicke model is given by
\begin{equation}
\begin{split}
	\hat H_\text{Dicke}&=\omega\hat a^\dagger \hat a+\omega_0\hat S_z+\frac{\lambda_-}{\sqrt{2S}}\left(\hat a\hat S_++ \hat a^\dagger\hat S_-\right)\\
	&+\frac{\lambda_+}{\sqrt{2S}}\left(\hat a\hat S_-+ \hat a^\dagger\hat S_+\right),
\end{split}
\end{equation}
expressed it in terms of collective spin operators  $\hat S_{\{x,y,z,\pm\}}=\sum_n^N \hat S^{(n)}_{\{x,y,z,\pm\}}$, where $\hat S^{(n)}_{\{x,y,z,\pm\}}$ are the spin operators of the $n$-th spinor.

In the limit $\sqrt{\omega^2+\kappa^2}\gg\omega_0,\lambda_\pm$, the cavity can be adiabatically eliminated to obtain \cite{masson2017}
\begin{equation}
\label{reducedmasterequation}
\begin{split}
    \dot{\hat\rho}&=-i\left[\hat H,\hat\rho\right]+\frac{\kappa({\mathfrak{n}}+1)}{2S(\omega^2+\kappa^2)}\mathcal{D}[\lambda_-\hat S_-+\lambda_+\hat S_+]\hat\rho\\
    &+\frac{\kappa{\mathfrak{n}}}{2S(\omega^2+\kappa^2)}\mathcal{D}[\lambda_-\hat S_++\lambda_+\hat S_-]\hat\rho ,
\end{split}
\end{equation}
with the Hamiltonian
\begin{equation}
\begin{split}
\label{Hamiltonianadbelim}
&\hat H=\left[\omega_0-\frac{\omega(2{\mathfrak{n}}+1)(\lambda^2_--\lambda^2_+)}{2S(\omega^2+\kappa^2)}\right]\hat S_z\\
&-\frac{\omega}{2S(\omega^2+\kappa^2)}\left[(\lambda_-+\lambda_+)^2\hat S_x^2+(\lambda_--\lambda_+)^2\hat S_y^2\right].
\end{split}
\end{equation}

We are considering a scheme, completely dissipative in nature, in particular, we will be setting $\omega=\omega_0=0$ in Eqs.~(\ref{reducedmasterequation}-\ref{Hamiltonianadbelim}). Additionally, we set $\lambda_+=-\lambda_-=-\lambda$. Both these changes can be achieved by tuning the frequency, strengths and phases of the driving and cavity fields \cite{groiseau2021,masson2017,dimer2007}. We end up with the master equation
\begin{equation}\label{evo}
    \dot{\hat\rho}=\Upsilon\mathcal{D}[\hat S_y]\hat\rho,
\end{equation}
where $\Upsilon=2\lambda^2(2\mathfrak{n}+1)/(S\kappa)$.


\section{Dissipative Freezing and Entangled-State Cycles}
Let us assume that our initial state is the Dicke state with magnetic quantum number $m=0$ in the $x$-basis (denoted by the subscript on the ket)
\begin{equation}
    |\psi(t=0)\rangle=|S,0\rangle_x.
\end{equation}
Since the evolution in Eq.~(\ref{evo}) depends only on $\hat S_y$, a change of basis from the eigenstates of $\hat S_x$ to those of $\hat S_y$ is in order. To this end we can use
\begin{equation}
\label{basis}
	|S,m'\rangle_x=\sum_{m=-S}^{S}d_{m,m'}^S\left(-\frac{\pi}{2}\right)|S,m\rangle_y ,
\end{equation}
where $d_{m,m'}^S(\beta )={}_z\langle S,m|e^{-i\beta\hat S_y}|S,m'\rangle_z$ are the elements of the Wigner $d$ matrix \cite{sakurai2020}. In the special case where $m'=0$, we can represent it in terms of the spherical harmonics $Y^m_S(\theta,\varphi)$, i.e.,
\begin{equation}
    d^S_{m,0}=\sqrt{\frac{4\pi}{2S+1}}Y^m_S(-\frac{\pi}{2},0).
\end{equation}
Note that if $S$ is even/odd, then only the even/odd Dicke states $|S,m\rangle_y$ have non-zero amplitudes $d^S_{m,0}$ and we have $d^S_{-m,0}=(-1)^Sd^S_{m,0}$.

We unravel the Master equation to describe the evolution of a single quantum trajectory \cite{carmichael1993}. Each time step $\Delta t$, we evolve the wave function with the non-Hermitian Hamiltonian $\hat H_{NH}=-i\Upsilon\hat S_y^2$. This comes with a loss proportional to the probability of a quantum jump, i.e., $\Delta t\langle\psi(t)|2\Upsilon\hat S_y^2|\psi(t)\rangle$. We determine whether a jump happened with a random number generator and if so apply $\hat S_y$ to the wave function. In both cases, we renormalize the wave function afterwards. With this we can rewrite the wave function between jumps with Eq.~(\ref{evo}) as
\begin{equation}
\begin{split}
    |\psi(t)\rangle&=\frac{e^{-\Upsilon\hat S_y^2 t}|\psi(0)\rangle}{||e^{-\Upsilon\hat S_y^2 t}|\psi(0)\rangle||}=\frac{\sum_m d^S_{m,0}e^{-\Upsilon m^2t}|S,m\rangle_y}{\sqrt{\sum_m e^{-2\Upsilon m^2t}(d^S_{m,0})^2}}\\
    &=\frac{d^S_{0,0}|S,0\rangle_y+\sum_{m=1}^S d^S_{m,0}e^{-\Upsilon m^2t}\sqrt{2}|\chi_S^+(m)\rangle}{\sqrt{\sum_m e^{-2\Upsilon m^2t}(d^S_{m,0})^2}},
\end{split}    
\end{equation}
where $|\chi_S^\pm (m)\rangle$
are the Kitten states
\begin{equation}
    |\chi_S^\pm(m)\rangle=\frac{1}{\sqrt{2}}\left(|S,m\rangle_y\pm(-1)^S|S,-m\rangle_y\right).
\end{equation}

\begin{figure}[H]
\centering
	\includegraphics[width=\linewidth]{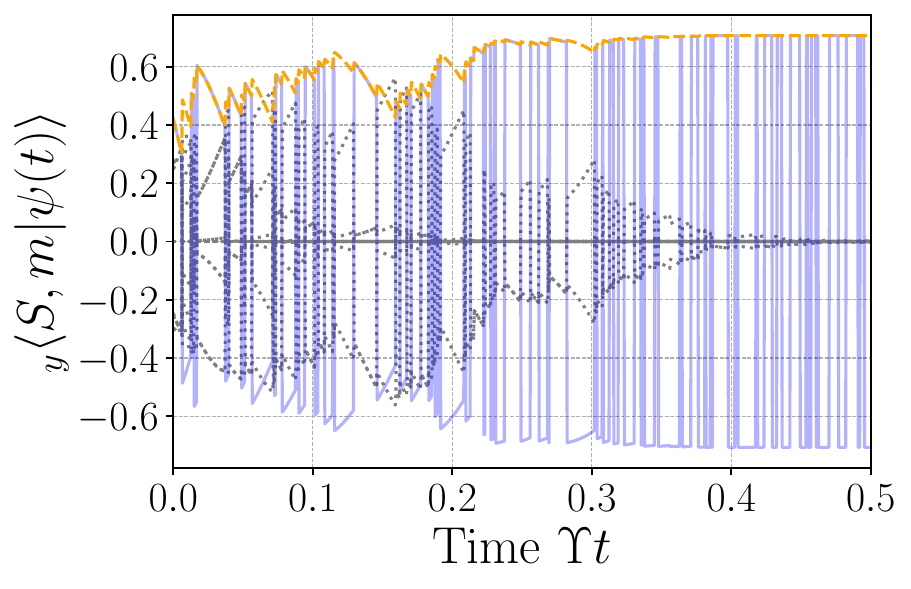}
	\caption{Single-trajectory simulation of the overlap of the system state with the eigenstates of $\hat S_y$. We plot the overlap with the two Dicke states $\langle\psi(t)|S,-m\rangle_y$ (solid blue) and $\langle\psi(t)|S,m\rangle_x$ (dashed orange) for the $m$-value that corresponds to the final, surviving entangled-state cycle, and the overlaps remaining Dicke states (dotted grey) as a function of time for a single trajectory of spin $S=10$. The trajectory shows a cycle that settles in the Cat state $m=S$. }
	\label{cycle}
\end{figure}

After the first jump (photoemission), the $|S,0\rangle_y$-component of $|\psi(t)\rangle$ is eliminated. Since all $|S,m\neq0\rangle_y$ decay, the system emits photons ad infinitum. FIG.~\ref{cycle} shows how in the limit of large time $t$ only one pair $|\chi_S^\pm(m)\rangle$ survives, due to the positive feedback loop between the rate of photoemissions and the populations of $|S,m\rangle_y$ (cf. \cite{groiseau2021,chia2008}). This can also be understood by $\hat S_y$ being a strong symmetry of the system (commuting with the Hamiltonian and the Lindblad operator) \cite{sanchezmunoz2019}. The probability for a specific entangled-state cycle to occur $P_{|\chi_S(m)\rangle}=2(d^S_{m,0})^2$ (without the factor 2 if $m=0$) is tantamount to the initial population $|{}_x\langle S,0|\chi_S^\pm(S)\rangle|^2$ because Dicke states $|S,m\rangle_y$ are steady states of Eq.~(\ref{evo}) in the trajectory average. In \cite{groiseau2021a} ideas to steer lower entangled-state cycles into higher ones were explored, but remained mostly hypothetical. Each jump also changes the relative sign within the Kitten states since
\begin{equation}
\label{jumpoperator}
    \hat S_y|\chi_S^\pm(m)\rangle=m|\chi_S^\mp(m)\rangle,
\end{equation}
so that the system switches back and forth between the two states $|\chi_S^\pm(m)\rangle$ (vertical lines in FIG.~\ref{cycle}). The rate at which it does this allows us to determine in what cycle the system has settled in, because the rate scales quadratically with $m$, i.e.,
\begin{equation}
    \langle\chi_S^\pm(m)|\hat S_y^2|\chi_S^\pm(m)\rangle=m^2 .
\end{equation}
The rate could be determined by performing a photon counting measurement on the cavity output. As highlighted in \cite{chia2008}, conducting a homodyne measurement on the cavity output instead of photon counting and thereby determining $\langle m\rangle$ instead of $\langle m^2\rangle$, could be used to force the system to settle into the corresponding Dicke state.

\begin{figure}[b]
\centering
	\includegraphics[width=\linewidth]{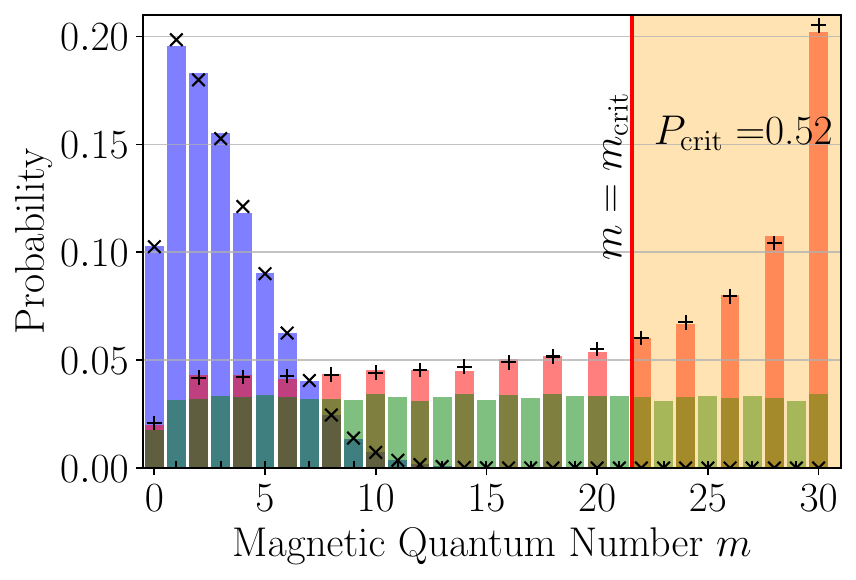}
	\caption{Histogram of the probability to land in each entangled-state cycle for the initial state $|S,S\rangle_x$ (blue) equal distribution of all $|S,0\rangle_y$ (green) and $|S,0\rangle_x$ (red), for $S=30$ and averaged over 50000 quantum trajectories. The vertical line represent the value of $m_\mathrm{crit}$, left of which lies the highlighted are that represents the entangled-cycles with increased QFI compared to $|S,0\rangle_x$ (with total probability $P_\text{crit}$). The black pluses and crosses represent the analytical values for the initial populations $|{}_x\langle S,0|\chi_S^\pm(S)\rangle|^2$ and $|{}_x\langle S,S|\chi_S^\pm(S)\rangle|^2$, respectively.}
	\label{Histogram}
\end{figure}

FIG.~\ref{Histogram} shows a histogram of the probability of ending in the entangled-state cycle $|\chi_S^\pm(m)\rangle$. The initial state $|S,0\rangle_x$ offers a distribution that is significantly skewed towards larger $m$, and is therefore much more likely to produce large Kitten states than a spin-coherent state \cite{groiseau2021,chia2008} or a uniform distribution \cite{sanchezmunoz2019}. In fact the Cat states $|\chi_S^\pm(S)\rangle$ are the most likely outcome with probability
\begin{equation}
    P_\text{Cat}= 2(d^S_{S,0})^2=2\frac{(2S)!}{2^{2S}(S!)^2}\approx\frac{2}{\sqrt{\pi S}},
\end{equation}
which is a factor $\exp(S)$ larger than if we had started with a spin-coherent state \cite{chia2008,groiseau2021}.
Here we used that $Y_S^{\pm S}(\theta,0)=(\mp1)^S\sqrt{(2S+1)!/(4\pi)} \sin^S\theta /(2^S S!)$ and Stirling's formula ($S!\approx S^S e^{-S}\sqrt{2\pi S}$).

\section{Potential Applications}

\subsection{Quantum Secret Sharing}

In quantum secret sharing, the goal is that sender Alice $A$ shares a secret with a number of other parties called Bob $B$ so that only together all Bobs can access the secret \cite{hillery1999}. The Kitten states in the standard basis representation ($N=2S$ spin $1/2$-spinors) are given by
\begin{equation}
|\chi^\pm_S(m)\rangle= \frac{\sum( \left|1^{\otimes S+m} 0^{\otimes S-m}\right\rangle_y  \pm \left|1^{\otimes S-m}0^{\otimes S+m}\right\rangle_y)}{\sqrt{2\binom{2S}{S-m}}},
\end{equation}
where we sum all permutations. The crucial property for quantum secret sharing is that
\begin{equation}
\begin{split}
\langle\chi^\pm_S(m)|\hat\sigma_z^{\otimes2S}|\chi^\pm_S(m)\rangle&=\pm 1,\\
\langle\chi^\pm_S(m)|\hat\sigma_z^{\otimes2S}|\chi^\pm_S(m)\rangle&=\pm(-1)^{S+ m}.
\end{split}
\end{equation}
This allows the $2S-1$ Bobs if and only if they perform the same measurement with the results $R_{B_i}$ and cooperate to deduce the result of Alice's measurement result $R_A$ via $R_A=\langle\hat\sigma_{z,x}^{\otimes2S}\rangle/(R_{B_1}...R_{B_{2S-1}})$. As soon as one Bob is taken out of this process, Alice's result is left secret. The scheme does require exact knowledge of the initial state, which would require keeping track of the number of photodetections during the generation of the Kitten state.

An eavesdropper Evan with an associated ancillary qubit $E$ might spoil the secret. 
We have
\begin{equation}
    |\chi^\pm_S(m)\rangle=\frac{1}{\sqrt{2}}(|0\rangle_A|\xi\rangle_B+|1\rangle_A|\overline{\xi}\rangle_B),
\end{equation}
where
\begin{equation}
    \begin{split}
        \left|\xi\right\rangle &= \frac{\sum( \left|1^{\otimes S+m} 0^{\otimes S-m-1}\right\rangle_y  \pm \left|1^{\otimes S-m}0^{\otimes S+m-1}\right\rangle_y)}{\sqrt{2\binom{2S}{S-m}}}\\
\left|\overline{\xi}\right\rangle &= \frac{\sum(\left|1^{\otimes S+m-1} 0^{\otimes S-m}\right\rangle_y  \pm \left|1^{\otimes S-m-1}0^{\otimes S+m}\right\rangle_y)}{\sqrt{2\binom{2S}{S-m}}},
    \end{split}
\end{equation}
where the $\pm$ relates to $|\chi^\pm_S(m)\rangle$. The optimal coherent individual attack tries to establish entanglement between Alice and Evan via
\begin{equation}
\begin{split}
    \hat U_{BE} \left|\xi\right\rangle_B\left|0\right\rangle_E &=      \left|\xi\right\rangle_B\left|0\right\rangle_E\\ 
\hat U_{BE} \left|\overline{\xi}\right\rangle_B\left|0\right\rangle_E & =  \cos \phi \left|\overline{\xi}\right\rangle_B\left|0\right\rangle_E+  \sin \phi \left|\xi\right\rangle_B\left|1\right\rangle_E,
\end{split}
\label{unitary}
\end{equation} 
where $\phi\in[0,\pi/2)$ corresponds to the strength of the attack. This leaves the system in the quantum state
\begin{equation}
\begin{split}
       &\hat U_{BE}|\chi^\pm_S(m)\rangle|0\rangle_E=\frac{1}{\sqrt{2}}(|0\rangle_A|\xi\rangle_B|0\rangle_E\\
       &+\cos\phi|1\rangle_A|\overline{\xi}\rangle_B|0\rangle_E+\sin\phi|1\rangle_A|\xi\rangle_B|1\rangle_E),
\end{split}
\end{equation}
which takes the same form as for a $G$-state ($|\chi^\pm_S(S-1)\rangle$). This was shown to be secure in the regime $\phi\in[0,\pi/4)$, i.e. the two-qubit Bell inequality is violated between Alice and any of the Bobs and not between Alice and Evan \cite{sende2003}.

\subsection{Quantum Metrology}
\begin{figure}[tbh]
\centering
	\includegraphics[width=0.8\linewidth]{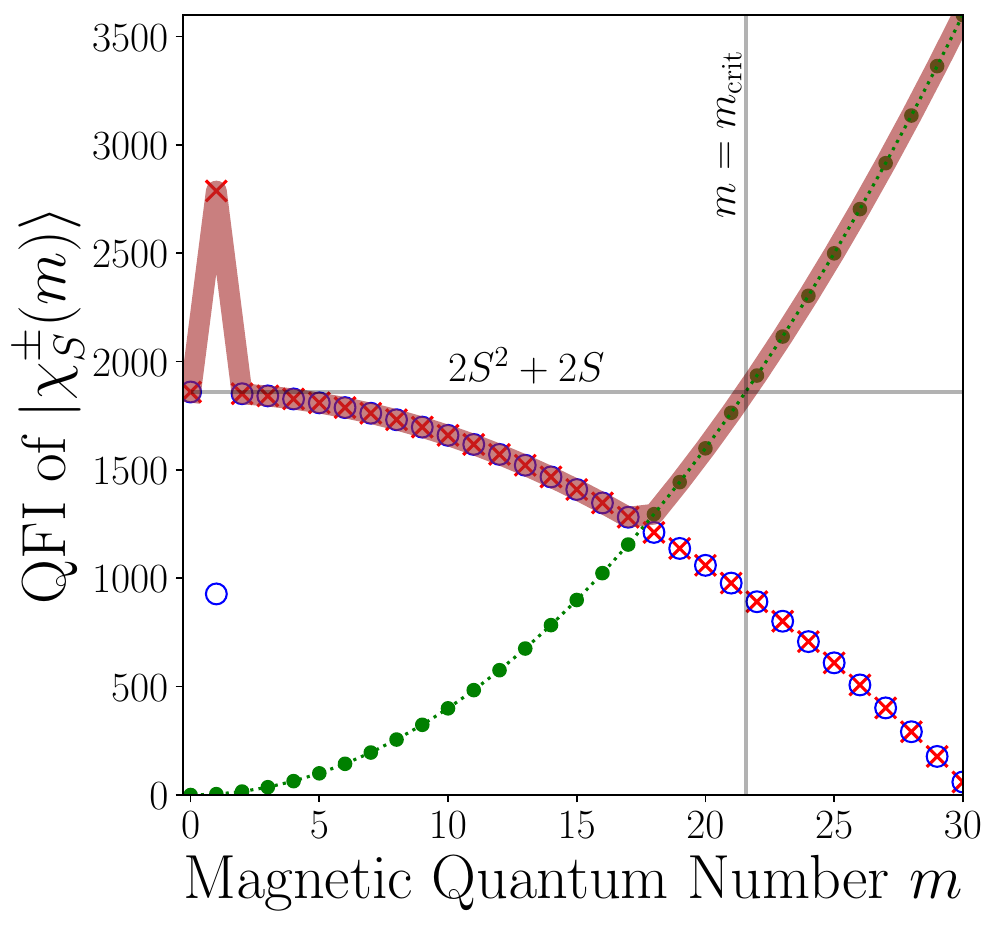}
	\caption{QFI of the state $|\chi^\pm_S(m)\rangle$ with respect to the generators $\hat S_x$ (blue circles), $\hat S_y$ (green dots) and $\hat S_z$ (red crosses) for $S=30$. The broad brown curve shows the QFI with the optimal generator from all generators of the form $\sin\theta\cos\varphi\hat S_x+\sin\theta\sin\varphi\hat S_y+\cos\theta\hat S_z$. For $m=0$ the state is just $|S,0 \rangle_y$.}
	\label{QFI}
\end{figure}

The sensitivity of a pure state $|\psi\rangle$ exposed to a unitary transformation generated by the Hermitian operator $\hat O$ can be quantified with the Quantum Fisher Information (QFI) and is related to its variance, i.e., QFI~$=4(\langle\psi|\hat O^2|\psi\rangle-\langle\psi|\hat O|\psi\rangle^2)$ \cite{braunstein1994}. The Kitten states $ |\chi_S^\pm(m)\rangle$ have QFIs equal to $4m^2$, with respect to the generator $\hat S_y$. Note that for the small Kitten states that is not the optimal choice for the QFI ($\hat S_x$ or $\hat S_z$ are better generators). This is shown in FIG. \ref{QFI}, where we plot the QFI with respect to all linear combinations of $\hat S_x,\hat S_y$ and $\hat S_z$. This can also be seen by imagining tracing orthodromes/great circles on the Wigner functions in FIG.~\ref{idea} and looking for the one that cuts through the sharpest/deepest interference fringes (the corresponding generator will be orthogonal to the plane spanned by the great circle). For the small Kitten state, for example, the interference pattern along the equator is very coarse but along the meridian almost as fine as for the Dicke state. Compared to $|S,0\rangle_x$ with QFI~$=2S^2+2S$ (with respect to $\hat S_z,\hat S_y$) we can define a lower threshold for the magnetic quantum number $ m_\text{crit}$ beyond which the QFI is larger than that of $|S,0\rangle_x$, i.e.,
\begin{equation}
    m_\text{crit}=\sqrt{\frac{S(S+1)}{2}}\approx\frac{S}{\sqrt{2}}.
\end{equation}
The probability of an improvement in QFI is therefore the cumulative probability of all entangled-state cycles above that threshold (highlighted region in FIG.~\ref{Histogram})
\begin{equation}
\begin{split}
        P_\text{crit}&=\sum_{m=\lceil m_\text{crit}  \rceil}^S 2(d^S_{m,0})^2=\sum_{b=0}^{S-\lceil m_\text{crit}  \rceil} 2(d^S_{S-b,0})^2,
\end{split}
\end{equation}
where $\lceil x  \rceil$, $\lfloor x \rfloor$ are the ceiling and floor functions.

\begin{figure}[tbh]
\centering
	\includegraphics[width=\linewidth]{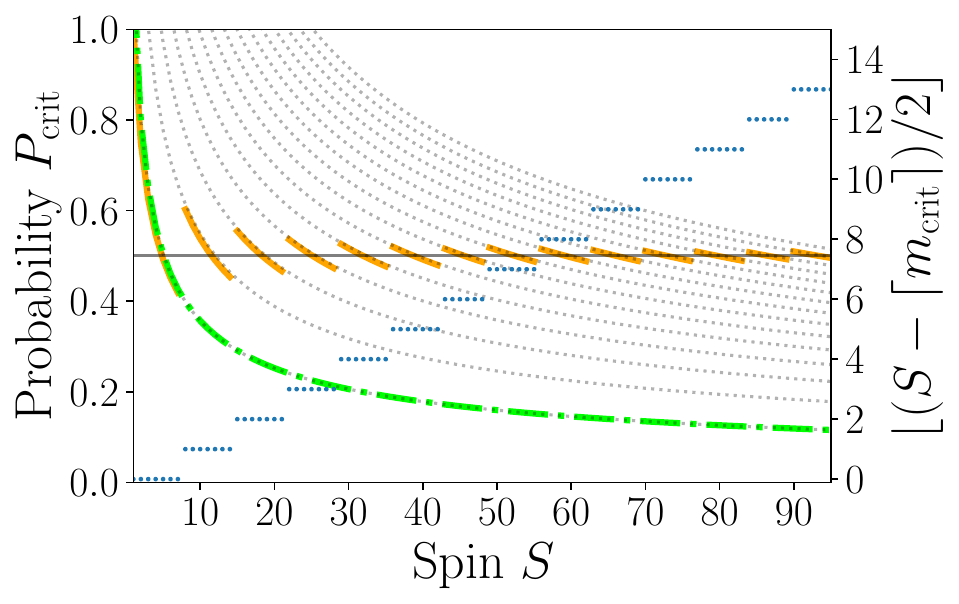}
	\caption{Plot of the probability to increase the QFI $P_\text{crit}$ (solid orange), its piecewise approximation $2\alpha_r/\sqrt{\pi S}$ (dotted gray), $P_\text{Cat}$ (dashdotted green). Note that, $P_\text{Cat}=2\alpha_0/\sqrt{\pi S}$. The other axis shows the number of non-zero population magnetic quantum number $m$ that contribute to $P_\text{crit}$, i.e., $\lfloor (S-\lceil m_\text{crit}\rceil)/2\rfloor$ (blue dots).}
	\label{Pcrit}
\end{figure}

FIG.~\ref{Pcrit} shows that $P_\text{crit}$ becomes discontinuous when an additional magnetic quantum number $m$ with non-zero population becomes larger than $m_\mathrm{crit}$ and starts contributing to $P_\text{crit}$ (whenever $S-\lceil m_\text{crit}\rceil$ increases by two or equivalently $\lfloor (S-\lceil m_\text{crit}\rceil)/2\rfloor$ by one, because only every second $m$ has non-zero population). In the limit of large $S$, we can see that $\lfloor (S-\lceil m_\text{crit}  \rceil)/2\rfloor\xrightarrow[S\to \infty]{} \frac{\sqrt{2}-1}{2\sqrt{2}}S\approx  S/6.82$, which means that the $r$-th jump happens when the spin becomes equal to $S_r=1+\frac{2\sqrt{2}}{\sqrt{2}-1}r\approx1+7r$ (this periodicity of 7 gets occasionally interrupted by a segment of length 6, see FIG.~\ref{Pcrit} between the sixth and seventh jump). This means we can express $P_\text{crit}$ piecewise on the segment $S\in[S_r,S_{r+1})$ as
\begin{equation}
\begin{split}
               P_\text{crit} &=2(d^S_{S,0})^2\alpha_{r} \approx \frac{2\alpha_{r}}{\sqrt{\pi S}},
\end{split}
\end{equation}
where we defined the spin-dependent prefactors
\begin{equation}
\begin{split}
        &\alpha_r(S)=\sum_{b=0}^r \left[\frac{d^S_{S-2b,0}}{d^S_{S,0}}\right]^2\\
        &\approx1+\sum_{b=1}^r\sqrt{\frac{S}{\pi b(S-b)}}\frac{1+\frac{1}{24b}}{(1+\frac{1}{12b})^2}.
\end{split}
\end{equation}
The spin dependence ends up being negligible over the segment they are defined on, meaning that any $S\in[S_r,S_{r+1})$ will yield roughly the same $\alpha_r(S)$. For a detailed explanation of this result, see Appendix \ref{calculations}. Therein it is also shown that $P_\text{crit}$ approaches $1/2$ for large $S$ which supports the numerical results in FIG.~\ref{Pcrit}.


\section{Experimental Feasibility}
The feasibility of this scheme in practice depends on two key factors: the ease of obtaining the initial state $|S,0\rangle_x$ and how well the evolution can be shielded against noise.

\subsection{The Initial State}

The Dicke state $|S,0\rangle$ possesses strong spin squeezing and has been studied extensively \cite{raghavan2001,luo2017,thiel2007,zou2018,masson2019,stockton2004,lucke2014}. It should be noted that the initial state $|S,0\rangle_x$ can also be generated from the same model in Eqs.~(\ref{reducedmasterequation}-\ref{Hamiltonianadbelim}) as was detailed in \cite{chia2008,groiseau2021}. For this, we start from the initial Dicke state $|S,S\rangle_z$ (easily achieved by pumping all spinors to their highest state) and set $\lambda_+=\lambda_-=\lambda$, leading to the master equation $\dot{\hat\rho}=\Upsilon\mathcal{D}[\hat S_x]\hat\rho$. The system settling in $|S,0\rangle_x$ (with probability $P\approx1/\sqrt{\pi S}$), is heralded by the lack of any photoemissions and can therefore be postselected for.  In \cite{masson2019}, $|S,0\rangle$ was theoretically generated with an efficiency of $15.5\%$  and a fidelity of $99.52\%$ for $N=100$ spin $1/2$ particles. In \cite{zou2018}, these states are experimentally generated deterministically in ensembles of more than $10^4$ spin-1 particles, but the final state is a superposition of Dicke states with lower total spin, i.e., $\sum_{S'}|S',0\rangle$. This would not significantly decrease the resulting QFI, as long as the distribution of $S'$'s is not too broad. The reported distributions are quite narrow with ranges of $0.99S<S'<S$. If inserted into our scheme this would result in superposition of different entangled-state cycles [similar $m$ (same if only one parity involved, most likely one parity will dominate over time) but different $S'$].

\subsection{Single-particle Decoherence}

\begin{figure}[tbh]
\centering
 \includegraphics[width=\linewidth]{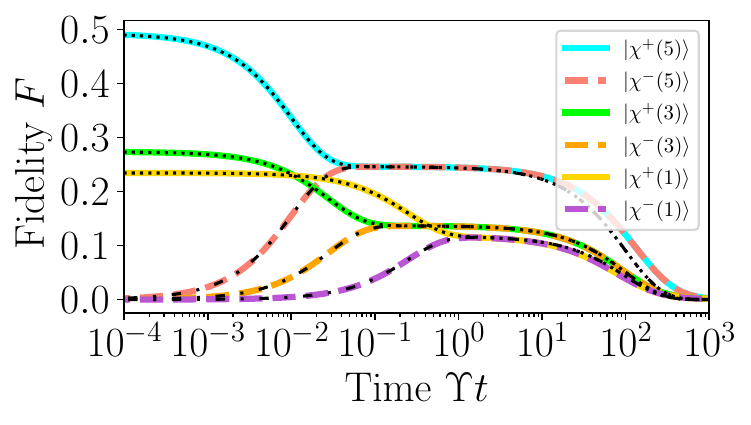}
	\caption{Simulations of single-particle noise using permutational invariance starting from $|S,0\rangle_x$. We plot the fidelity $F=\left(\text{Tr}\sqrt{\sqrt{\hat \rho(t)}|\chi^\pm(m)\rangle\langle\chi^\pm(m)|\sqrt{\hat \rho(t)}}\right)^2$ of the quantum state during the evolution with the even (colored solid lines) and odd Kitten states (colored dashed lines) and the analytical fits $(d_{m,0}^S)^2\exp(-\gamma_\text{eff}Nt)[1-\exp(-4\Gamma m^2 t)]$ (black dashdotted line) and $(d_{m,0}^S)^2[\exp(-\gamma_\text{eff}Nt)+\exp(-4\Gamma m^2 t)]$ (black dotted line). The parameters are $S=5,N=10$ and $\gamma_\text{eff}/\Upsilon=0.001$.}
	\label{timescale}
\end{figure}

We use a general model for single-particle decoherence effects by including dephasing, pumping and decay of the individual atoms in the spin basis, which can be described with the Linbladians $\sum_n^N\left(\mathcal{D}[\hat S^{(n)}_z/2]\hat\rho+\mathcal{D}[\hat S^{(n)}_+]\hat\rho+\mathcal{D}[\hat S^{(n)}_-]\hat\rho\right)\gamma_{\rm eff}/2$, where for simulation purposes we assume all effects to be of equal strength $\gamma_{\rm eff}$ and the individual spinors to be of spin $1/2$, i.e., $S^{(n)}_i$ to be the Pauli matrices. We simulate these single-particle effects using the permutational quantum invariant solver (PIQS) \cite{shammah2018}. FIG.~\ref{timescale} shows the population of the different Kitten states over time under the influence of large collective and small single-particle decoherence. The moment an entangled-state cycle is reached in a single trajectory corresponds to reaching the steady state of the collective decay in the ensemble picture, i.e., when the population of the odd and even Kitten states with identical $m$ become equal. The timescale for this process is governed by the decay of the smallest Kitten state, which is given by $1/\Upsilon$. After that, we are entering the time scale of the single-particle noise $1/(\gamma_\text{eff}N)$, which depopulates the Kitten states. To be able to prepare our desired state before noise gets the better of it, we need $\Upsilon>\gamma_\text{eff}N$.

 One physical example for such processes could be the spontaneous emission from off-resonant excitation. For example, in the spin-1 implementation in ${}^{87}$Rb using the $D_1$-line \cite{masson2017,groiseau2021}, this means that $\Upsilon\simeq(2\mathfrak{n}+1)g^2 \Omega^2/(72\kappa\Delta^2)$
has to be larger than the effective decay rate from spontaneous emission given by $\gamma_\mathrm{eff} \simeq \gamma\Omega^2/(12\Delta^2)$,
where $g$ is the atom-cavity coupling, $\Omega$ is the Rabi frequency of the laser, $\Delta$ is the detuning between atom and laser/cavity and $\gamma$ is the natural line width of the $5^2P_{1/2}$ manifold. For $\mathfrak{n}=0$, the ratio of these rates is $\gamma_{\rm eff}N/\Upsilon\simeq 12 N/C$, where $C=2g^2/(\kappa\gamma)$ is the single-atom cooperativity. If we considered
$\gamma_{\rm eff}N/\Upsilon =5\times 10^{-2}$, it would require $C=240N$, which will be hard to achieve for large atomic ensembles. This requirement can be relaxed by a factor $2\mathfrak{n}+1$ if the cavity is thermally populated.

In principle, the idea can be experimentally implemented in an engineered Dicke model, especially cavity QED implementations seem suitable \cite{davis2019,zhiqiang2017} due to being able to monitor the cavity output. Depending on the atomic species, the spontaneous emission rates of alkaline atoms are typically of the order of $\gamma=3$-$6$ MHz. The experimental cavity parameters of current Dicke models \cite{davis2019,zhiqiang2017} have cooperativities below $C=10$, which would require very large $\mathfrak{n}$. Modern micro- and nanocavity experiments can reach cooperativities beyond $C=100$ \cite{sames2014,reimann2015,haas2014,gehr2010,barontini2015,thompson2013,tiecke2014,samutpraphoot2020}, so that setups with $N\approx \mathfrak{n}$ are imaginable, but $\mathfrak{n}$ is typically small.

It is therefore desirable to drastically increase the thermal population of the cavity. Doing that by increasing the temperature would require unrealistically high temperatures because of the large optical frequency and would also amplify spontaneous emission or other decoherence processes. A monochromatic thermal drive whose frequency matches that of the cavity as an input field could circumvent this issue. This could be achieved by sending black-body radiation through a monochromator or by using acousto-optical modulation of the output of a superluminescent diode \cite{straka2018}. Considering a two-sided cavity that interacts on one side with this hot reservoir (in) and on the other side with a reservoir in the vacuum state (out) would replace the current Lindbladform with $\kappa_\text{out}\mathcal{D}[\hat a]+\kappa_\text{in}(n_{th}+1) \mathcal{D}[\hat a]\hat\rho+\kappa_\text{in}n_{th} \mathcal{D}[\hat a^\dagger]\hat\rho$. Here, $\kappa_\text{in}$ is derived from the cavity-reservoir coupling on the input side and $\kappa_\text{out}$ on the other side. The resulting thermal cavity population would be $\mathfrak{n}=\kappa_\text{in}n_{th}/(\kappa_\text{in}+\kappa_\text{out})$, where $n_{th}$ is the population of the input thermal reservoir \cite{henninarencibia2023}. This would change the collective decay to $\Upsilon=2\lambda^2(2\mathfrak{n}+1)/[S(\kappa_\text{in}+\kappa_\text{out})]$.

Microwave-implementations of the engineered Dicke model have also been considered in \cite{grimsmo2014}, for example in fluxonium coupled to superconducting microwave cavities (frequencies of several GHz), the situation is similar with cooperativities $C>100$ \cite{earnest2018,lin2018}, and $\mathfrak{n}$ is naturally bigger.

Some states with weak multi-qubit entanglement end up being more robust against white noise than state with strong multi-qubit entanglement like GHZ states \cite{sende2003,laskowski2014}. Let us assume that we have such an admixture of our Kitten state (again for $N=2S$ spin $1/2$-spinors)
\begin{equation}
\hat\rho_{2S}=p|\chi^\pm_S(m)\rangle\langle\chi^\pm_S(m)|+(1-p)\mathbb{I}_{2S}/2^{2S}.
\end{equation}
If $2S-2$ of the parties sharing the state locally measure $\hat\sigma_y$ and the result is that $S+m-1$ parties measure 0 and $S-m-1$ parties measure 1 (or vice versa), then the resulting state for the two remaining parties is
\begin{equation}
    \hat\rho_2=\frac{1}{1+\frac{(1-p)\binom{2S}{S-m}}{ 2^{2S-2}p}}|\Psi^+\rangle\langle \Psi^+|+\frac{1}{1+\frac{ 2^{2S-2}p}{(1-p)\binom{2S}{S-m}}}\mathbb{I}_2/4,
\end{equation}
where $|\Psi^+\rangle=(|01\rangle+|10\rangle)/\sqrt{2}$. $\hat\rho_2$ doesn't adhere to local realism if $[1+(1-p)\binom{2S}{S-m}/( 2^{2S-2}p)]^{-1}>1/\sqrt{2}$. As this cannot arise from local operations on states adhering to local realism, $\hat\rho_{2S}$ must violate local realism if $p$ exceeds the critical probability
\begin{equation}
    p^\text{Kitten}_\text{crit}=\frac{\binom{2S}{S-m}}{\binom{2S}{S-m}+(\sqrt{2}-1)2^{2S-2}},
\end{equation}
which increase with decreasing $m$ leading to the Kitten states always being more robust than our initial state. This result is slightly larger than the critical value of Dicke states from \cite{laskowski2014} due to being a superposition of two Dicke states.
We find that the critical value for the magnetic quantum number number of the Kitten state for which the Kitten state becomes more robust than a GHZ state ($p^\text{GHZ}_\text{crit}=2^{(1-2S)/2}$) roughly scales linearly with the spin ($[(S-m)/2S]_\text{crit}\approx0.22$).

\section{Conclusions}

We have shown that choosing Dicke states from a basis orthogonal to the dynamics as an initial state to dissipative one-axis twisting allows for the generation of large Kitten states.
By choosing the eigenstate with magnetic quantum number zero we get a specific superposition that is biased towards the large magnetic quantum numbers and has been experimentally generated with satisfying fidelity in the past.

We have shown how the resulting Kitten states can have improved characteristics when it comes to quantum metrology or quantum secret sharing.
The success can be monitored via the cavity output and we can post-select for it, however, even if in case of a failure the loss in QFI is likely not very big because all Kitten states are metrologically relevant.

Finally, we want to remark that the scheme could technically also be applied to the other Dicke states $|S,m\rangle_x$ with $0<m<S$. In that case, we have an initial QFI~$=2S^2+2S-2m^2$ resulting in $m_\text{crit}=\sqrt{[S(S+1)-m^2]/2}$. Compared to the initial state $|S,0\rangle_x$, $m_\text{crit}$ is lower for Dicke states with $m>0$. The probability of increasing the QFI $P_\text{crit}$ can then also be greater for some of the intermediate Dicke states.
However, the average magnetic quantum number $\langle |m|\rangle$ and $P_\text{Cat}$ are still the largest for $|S,0\rangle_x$, so the resulting Kitten states and QFI are still going to be larger on average for $|S,0\rangle_x$. It should be noted that the average resulting QFI will be the largest for the spin-coherent state because then the production of a small Kitten state with QFI close to $2S^2+2S$ is guaranteed. For half-integer spins, since $|S,0\rangle$ does not exist, one can simply take $|S,\pm\frac{1}{2}\rangle$, which yields similarly biased distributions.

Other schemes that generate spin Cat states that are mainly based on the evolution with a one-axis twisting Hamiltonian. Our scheme's dissipative nature gives it the potential to operate at much faster time scales thanks to the thermal drive which can be beneficial in systems with limited coherence time. 
    Our scheme settles in its entangled-state in a time $1/\Upsilon\geq72\kappa\Delta^2/(2g^2)\approx2$ $\mu$s based on the realistic parameter set $\{\kappa,\Delta,g,\Omega\}/2\pi=\{0.05,25,0.25,4\}$ GHz \cite{barontini2015,groiseau2021}. Current experiments have time scales of the order of ms \cite{yang2025} or tenths of $\mu$s \cite{chalopin2018}, which could be overcome with the additional factor $2\mathfrak{n}+1$. It also does not require precise temporal control of the system parameters besides keeping the electromagnetic fields constant unlike schemes based on machine optimization \cite{huang2022}.

\begin{acknowledgments} 
This work makes use of the Quantum Toolbox in Python (QuTiP) \cite{johansson2012,johansson2013}.
We thank Scott Parkins, Carlos Sánchez Muñoz and Sandro Wimberger for helpful discussions and comments.
\end{acknowledgments}

\bibliographystyle{apsrev4-1}
\bibliography{SpinCats.bib}

\clearpage

\appendix
\onecolumngrid

\section{Calculations}
\label{calculations}
We can reformulate the spherical harmonics as
\begin{equation}
\begin{split}
    Y_{S-2b}^S(-\frac{\pi}{2},0)&=(-1)^{S-2b}\sqrt{\frac{(2S+1)(2b)!}{4\pi(2S-2b)!}}P^{S-2b}_S(0)\\
    &=(-1)^{S-2b}\frac{\sqrt{2S+1}\sqrt{\Gamma(2b+1)}{}_2F_1(-S,S+1,-S+2b+1,\frac{1}{2})}{\sqrt{4\pi}\Gamma(1-S+2b)\sqrt{\Gamma(2S-2b+1)}},
    \end{split}
\end{equation}
where $P$ is the associated Legendre polynomial and ${}_2F_1$ the hypergeometric function \cite{abramowitz1948}, which we can express in terms of $\Gamma$-functions
\begin{equation}
\begin{split}
    {}_2F_1(-S,S+1,-S+2b+1,\frac{1}{2})
    &=\frac{2^{S-2b}\Gamma(1-S+2b)}{\Gamma(\frac{1}{2}-S+b)\Gamma(b+1)}.
    \end{split}
\end{equation}
This allows us to compute the coefficients
\begin{equation}
\begin{split}
        &\alpha_r(S)=\sum_{b=0}^r \left[\frac{d^S_{S-2b,0}}{d^S_{S,0}}\right]^2\\
        &=\sum_{b=0}^r \left[\frac{Y_{S-2b}^S(-\frac{\pi}{2},0)}{Y_{S}^S(-\frac{\pi}{2},0)}\right]^2\\
        &=\sum_{b=0}^r \left[\frac{\sqrt{\Gamma(2b+1)}\Gamma(\frac{1}{2}-S)\sqrt{\Gamma(2S+1)}}{\sqrt{\Gamma(2S-2b+1)}\Gamma(b+1)\Gamma(\frac{1}{2}+b-S)2^{2b}}\right]^2\\
        &\approx1+\sum_{b=1}^r\sqrt{\frac{S}{\pi b(S-b)}}\frac{1+\frac{1}{24b}}{(1+\frac{1}{12b})^2},
\end{split}
\end{equation}
where, in the third step we used that for non-negative integers $j$, $\Gamma(\frac{1}{2}-j)=(-4)^j j!\sqrt{\pi}/[(2j)!]$ \cite{abramowitz1948} and the Stirling approximation [we include an additional order for the terms solely depending on the potentially small $b$, i.e., $b!\approx b^b e^{-b}\sqrt{2\pi b}(1+\frac{1}{12b})$] \cite{abramowitz1948}.

\begin{figure}[tbh]
\centering
 \includegraphics[width=0.5\linewidth]{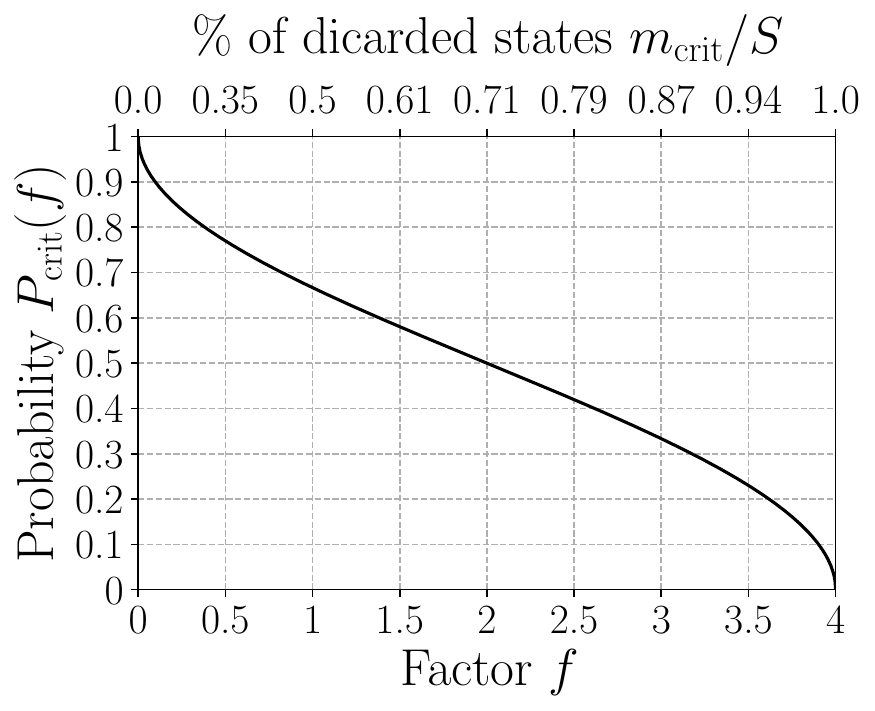}
	\caption{Plot of the probability $P_\text{crit}(f)$ to overcome a QFI of $fS^2$ against $f$ in the limit of infinite $S$. We also show what a given $f$ means in terms of the ratio of Kitten states that have to be overcome $m_\text{crit}/S=\sqrt{f}/2$.}
	\label{factor}
\end{figure}

Since the function in the series is integrable and decreasing monotonically for positive $b$, we can estimate a lower bound with the definite integral
\begin{equation}
    \alpha_r(S)\geq 1+\int_{1}^{r+1} db\sqrt{\frac{S}{\pi b(S-b)}}\frac{1+\frac{1}{24b}}{(1+\frac{1}{12b})^2},
\end{equation}
which we can then use to estimate the critical probability in the limit of large spin, i.e.,
\begin{equation}
\begin{split}
         P_\text{crit}&\geq  \frac{2[1+\int_1^{\lfloor (S-\lceil m_\text{crit}  \rceil)/2\rfloor+1}db\sqrt{\frac{S}{\pi b(S-b)}}\frac{1+\frac{1}{24b}}{(1+\frac{1}{12b})^2}]}{\sqrt{\pi S}}\\
         &=  \frac{2[1+\int_1^{\lfloor (S-\lceil m_\text{crit}  \rceil)/2\rfloor+1}db\sqrt{\frac{S}{\pi b(S-b)}}(1+\mathcal{O}(\frac{1}{b}))]}{\sqrt{\pi S}}\\
        &=2[\frac{1}{\sqrt{\pi S}}+\frac{2}{\pi}\arctan\sqrt{\frac{b}{S-b}}\Big \vert^{\lfloor \frac{S-\lceil m_\text{crit}  \rceil}{2}\rfloor+1}_1+\mathcal{O}(S^{-\frac{1}{2}})]\\
        &\xrightarrow[S\to \infty]{} \frac{4}{\pi}\arctan\sqrt{\frac{\sqrt{2}-1}{\sqrt{2}+1}}\\
        &=\frac{1}{2}.
\end{split}
\end{equation}
This can be generalized to the case, where we demand that the resulting QFI has to be larger than $fS^2$ ($ m_\text{crit}=fS^2$) with $f\in[0,4]$, i.e.,
\begin{equation}
    P_\text{crit}(f)=\frac{4}{\pi}\arctan\sqrt{\frac{2-\sqrt{f}}{2+\sqrt{f}}},
\end{equation}
which is depicted in FIG.~\ref{factor}.

\twocolumngrid

\end{document}